\documentstyle[12pt,tighten,aps]{revtex}
\draft

\begin{document}

\title{\bf Lyapunov instability of fluids composed of rigid diatomic molecules}

\author{Istv\'an Borzs\'ak}

\address{Laboratory of Theoretical Chemistry, E\"otv\"os University,\\
         H-1518 Budapest 112, Pf.32, Hungary}

\author{H. A. Posch}

\address{Institut f\"ur Experimentalphysik, Universit\"at Wien,\\
         Boltzmanngasse 5, A-1090 Wien, Austria}

\author{Andr\'as Baranyai}
\address{Laboratory of Theoretical Chemistry, E\"otv\"os University,\\
         H-1518 Budapest 112, Pf.32, Hungary}

\date{\today}

\maketitle

\begin{abstract}
   We  study  the Lyapunov instability of a two-dimensional
fluid composed of rigid diatomic molecules, with two
interaction sites each, and interacting with a WCA site-site
potential. We compute
full spectra of Lyapunov exponents
for such a molecular system. These exponents characterize
the rate at which neighboring trajectories diverge or converge
exponentially in phase space. Qualitative different degrees of
freedom -- such as rotation and translation  -- affect the Lyapunov
spectrum differently. We study this phenomenon by systematically varying
the molecular shape and the density. We define and evaluate ``rotation
numbers'' measuring the time averaged modulus of the
angular velocities for vectors connecting perturbed
satellite trajectories with an unperturbed reference trajectory in
phase space. For reasons of comparison, various time correlation functions
for translation and rotation are computed. The relative dynamics of
perturbed trajectories is also studied in certain subspaces of the
phase space associated with center-of-mass and orientational molecular motion.
\end{abstract}

% 05.45.+b Theory and models of chaotic systems
% 02.70.Ns  Molecular dynamics and particle methods
% 05.20.-y  Statistical mechanics
% 66.90.+r  Other topics in nonelectronoc transport properties
%           of condensed matter


\section{Introduction}

    The chaotic molecular motion in fluids and (nonlinear)
solids has been mainly studied in the past in terms of
correlation functions or related power spectra of assorted dynamical
variables. There is, however, a more fundamental point of view:
The basic underlying dynamical processes are collisions
(interactions) between particles with convex potential surfaces.
As a consequence, the phase space trajectory is highly unstable which is
reflected in a very sensitive dependence on initial conditions.
This phenomenon is characterized in terms of the set
of Lyapunov exponents $\{\lambda_l\}, l = 1,\cdots, L $, usually ordered
from the largest to the smallest. The
largest exponent $\lambda_1$ describes the time-averaged logarithmic
rate at which nearby phase-space trajectories separate. The sums
of exponents $\sum_{i=1}^{l} \lambda_i$ describe the expansion or
contraction rates of $l-$dimensional phase-space objects. Thus,
the Lyapunov exponents represent the time-averaged local deformation
rates in the neighborhood of a phase-space trajectory specified by the
time evolution of  specially-selected perturbation
vectors $\mbox{\boldmath $\delta$}_{l}(t)$. From a practical point of view
the precise orientation of the set of initial vectors
$\{ \mbox{\boldmath $\delta$}_{l}(0) \} $ is not known, nor is it needed for
the determination of the $\lambda_l$. This is discussed in more detail
in Section II.
The total number $L$ of exponents is
equal to the dimensionality of the phase space, and the whole set
of exponents is referred to as the Lyapunov spectrum.
For the evaluation of all $L$ exponents the simultaneous
integration of $ L \times (L+1) $ first-order differential
equations is required. The method has therefore been restricted to
rather low-dimensional dynamical systems in the past.
The feasibility of such studies for many-body systems has been demonstrated
by Hoover and Posch \cite{phpr88,phpr89}
who investigated fluid systems with up to 100 atoms in two dimensions.

   Lyapunov spectra of Hamiltonian systems in thermal equilibrium
exhibit a pronounced symmetry which helps to reduce the number of
differential equations to $L(L+2)/2$: for each positive exponent there exists
another negative exponent with equal absolute magnitude. This is
referred to as Smale pairing \cite{phpr88} or conjugate pairing
\cite{sem90,sem92}.  It is a consequence
of the symplectic nature of the equations of motion \cite{arnold},
which means that the phase flow -- viewed as a canonical
transformation of the phase space onto itself -- leaves the differential
two-form $\sum_{i=1}^{L/2} dp_{i} \wedge dq_{i}$
invariant. Here the sum is over all degrees of freedom, and $p_{i}$ and $q_{i}$
denote all components of particle momenta and positions, respectively.
Due to this pairing symmetry the calculation may be restricted to the
positive exponents, thus permitting the simulation of more complex
and larger systems.With the available computer hardware systems with
up to 400 degrees of freedom may be simulated at present
\cite{phpr94} by far exceeding the complexity one usually encounters
when studying dynamical systems with a low-dimensional phase space
\cite{hhp90}.

    For atomic fluids one finds that the shape of the
Lyapunov spectrum changes qualitatively if the density is isothermally
increased from that of a dense gas to solid densities \cite{phh90}, and that
the largest Lyapunov exponent $ \lambda_{1}$ exhibits
a maximum at the solid-fluid phase transition density
\cite{phh90},\cite{dellago1}. From a simulation of thermostatted
butane molecules one may further deduce that the largest contribution
to $ \lambda_{1}$ - and therefore the largest source for chaos - is
due to the torsional motion around the central CC-bond. The other
degrees of freedom contribute mainly to the smaller exponents \cite{toxpo}.
This conclusion has been reached by observing the variations
in the Lyapunov spectrum
of a Nos\'e--Hoover-thermostatted molecule, if various contributions
to the molecular Hamiltonian specifying different degrees of freedom
are selectively switched off.
Considering these examples we may expect that also for molecular fluids
the stretching and contraction properties are very differently affected by
translation and rotation, and that - as a consequence - the shape
of the Lyapunov spectra will strongly depend on the state of the system.
To clarify this point we report in this paper first results of a
molecular dynamics simulation study of a
simple planar molecular fluid in equilibrium consisting of $N$ rigid
homonuclear diatomic molecules with anisotropy $d/\sigma$. Here, $d$
is the fixed bond length separating the two interaction sites of a
molecule, and $\sigma$ is an atomic-size parameter
$d$, or the related moment of inertia,
and the molecular number density $n^{*}$ are two independent parameters
with which the emphasis may be conveniently shifted between
translational and rotational dynamics, at the same time varying
the respective relaxation times. We measure
the associated changes in the (in)stability properties of
the phase space trajectories.

    In Section II we define our model fluid and indicate our
method of simulating the reference trajectory. In Section III we
shortly outline the methods for the evaluation of Lyapunov spectra.
We define so-called ``rotation spectra'' for the
vectors $\mbox{\boldmath $\delta$}_l$ connecting neighboring trajectories
in phase space \cite{phh90}. It turns out that it is
useful to study also the projected motion of these vectors in subspaces of the
phase space such as the respective configurational and momentum spaces
for translational and rotational motion. To make contact with the
more traditional analysis of molecular dynamics in fluids, we also
evaluate correlation functions pertinent for these degrees of freedom.
In Section IV we summarize our results. They are discussed
further in the concluding Section V, where we speculate also about
possible interpretations of the Lyapunov spectra
in terms of suitably defined collective modes in dense fluids.

\section{The model fluid}

    Simulations were performed for a purely classical system consisting of
$N = 18$  rigid homonuclear diatomic molecules in a two-dimensional
square box of area (volume) $V$ and
with periodic boundary conditions.  The two interaction sites
of each molecule are separated by a rigid distance $d$ along the molecular
axis, and sites on different molecules separated by $r$ interact
with a purely repulsive Weeks--Chandler--Anderson potential,
\begin{equation}
\phi(r) = \left\{
\begin{array}{ccl}
4 \epsilon [(\sigma/r)^{12} - (\sigma/r)^{6} ]
                  + \epsilon & , & r< 2^{1/6}\sigma  \\
0 & , & r \geq 2^{1/6}\sigma \end{array} \right.   .
\label{27}
\end{equation}
The total intermolecular potential energy $\Phi$ is taken to be
pairwise additive. Reduced units are used throughout, for which
$\epsilon$, $\sigma$ and the atomic mass $m$ are unity. The equations of
motion for the $2N$ interaction sites were augmented with
constraint forces keeping the
bond length $d$ for each molecule fixed \cite{smnl}:
\begin{equation}
\begin{array}{ccl}
\dot {\bf q}_{\alpha} & = & {\bf p}_{\alpha}/m \\
\dot {\bf p}_{1} & = & {\bf f}_{1}  + \mu ({\bf q}_{2} - {\bf q}_{1}) \\
\dot {\bf p}_{2} & = & {\bf f}_{2}  - \mu ({\bf q}_{2} - {\bf q}_{1}).
\end{array}
\label{24}
\end{equation}
Here, ${\bf q}_{\alpha}, {\bf p}_{\alpha}$ are the
respective position and momentum vectors of the two
interaction sites of a molecule, $ \alpha = 1,2$,
and  ${\bf f}_{\alpha} =  -{\bf \nabla}_{\alpha} \Phi $ is the
force on site $\alpha$. Furthermore,
\begin{equation}
\mu =  \frac{({\bf p}_{2} - {\bf p}_{1})^{2}/m + ({\bf q}_{2} - {\bf q}_{1})
                  \cdot ({\bf f}_{2} - {\bf f}_{1}) }
      { 2 ({\bf q}_{2} - {\bf q}_{1})^{2} }
\label{25}
\end{equation}
is the Lagrange multiplier for the holonomic constraint
$ ({\bf q}_{2} - {\bf q}_{1})^{2} - d^{2} = 0 $ for this molecule.
The initial conditions were chosen such that the total center
of mass velocity was zero.

     For the evaluation of the Lyapunov exponents and the ensuing discussion
it is useful to represent the state of the system  by the $6N$-dimensional
state vector ${\bf \Gamma}(t) =
\{x_{i}, y_{i}, \dot{x}_{i}, \dot{y}_{i}, \eta_{i}, \dot{\eta}_{i}\}$,
$i = 1,N$, where $x_{i},y_{i}$ and
$\dot{x}_{i}, \dot{y}_{i}$ are the center of mass coordinates and momenta,
respectively, of molecule $i$, and $\eta_{i}$ is the angle the molecular axis
of $i$ makes  with some arbitrary direction. $\dot{\eta}_{i}$ is
the corresponding angular velocity. For each molecule $i$
these quantities may be easily computed from the instantaneous
positions and momenta of the interaction sites.

\section{Lyapunov exponents and rotation numbers}

     The equation of motion for the state vector ${\bf \Gamma}(t)$
is conveniently written as an autonomous
system of first-order differential equations:
\begin{equation}
\dot{{\bf \Gamma}}(t) = {\bf G}({\bf \Gamma}(t)) .
\label{31}
\end{equation}
Its solution defines  a flow ${\bf \Gamma}(t) = \Phi_{t}({\bf \Gamma}(0))$
in phase space. Let ${\bf \Gamma}(0)$ be the initial condition of a
reference trajectory, ${\bf \Gamma}_{s}(0)$ the initial point of a
neighboring perturbed trajectory, and let these two points be
connected by a parametrized path $C_{0}(s)$ with a perturbation parameter
$s$ such that
$\lim_{s \to 0}{\bf \Gamma}_{s}(0) = {\bf \Gamma}(0)$. At time $t$ these
points will be mapped by the flow into the points
${\bf \Gamma}(t) = \Phi_{t}({\bf \Gamma}(0))$ and
${\bf \Gamma}_{s}(t) = \Phi_{t}({\bf \Gamma}_{s}(0))$, and the path $C_{0}(s)$
into $C_{t}(s)$. Now we can define a finite-length tangent vector at $t=0$,
\begin{equation}
\mbox{\boldmath $\delta$}(0) = \lim_{s \to 0} \frac{{\bf
     \Gamma}_{s}(0) - {\bf \Gamma}(0)}{s},
\label{32}
\end{equation}
associated with an initial perturbation
$ {\bf \Gamma}_{s}(0) - {\bf \Gamma}(0)$  of the reference
trajectory in phase space. As time goes on, this
perturbation develops into $ {\bf \Gamma}_{s}(t) - {\bf \Gamma}(t)$, and
the associated tangent  vector becomes
\begin{equation}
\mbox{\boldmath $\delta$}(t) =
     \lim_{s \to 0} \frac{{\bf \Gamma}_{s}(t) - {\bf \Gamma}(t)}{s}.
\label{33}
\end{equation}
The change of the length of this vector during the time interval $t$
determines the stability of the reference trajectory due to the
initial infinitesimal perturbation. $\mbox{\boldmath $\delta$}(t)$
may be viewed as a vector comoving and corotating with the phase flow
in the immediate neighborhood of the phase point. It specifies a
direction in phase space which varies with time.
The equations of motion for $\mbox{\boldmath $\delta$}(t)$ are obtained
by linearizing the original motion equations (\ref{31}),
\begin{equation}
\dot{\mbox{\boldmath $\delta$}}(t) = {\bf D}({\bf \Gamma}(t)) \cdot
      \mbox{\boldmath $\delta$}(t),
\label{34}
\end{equation}
where ${\bf D}({\bf \Gamma})\equiv [(\partial/\partial{\bf \Gamma}){\bf G}
({\bf \Gamma})]$ is an $L \times L$ matrix and is referred to as
the stability matrix.
The formal solution of (\ref{34}) may be written as
\begin{equation}
\mbox{\boldmath $\delta$}(t) = {\bf M}(t;\mbox{\boldmath $\delta$}(0))
       \cdot \mbox{\boldmath $\delta$}(0) ,
\label{35}
\end{equation}
where the operator ${\bf M}$ is a time ordered
exponential of $\int{\bf D}((\Gamma(t'))dt'$.
According to the multiplicative ergodic theorem by Oseledec
\cite{oseledec,erue} the Lyapunov exponents
\begin{eqnarray}
\lambda_{l} & = &  \lim_{t \to \infty} \frac{1}{t} \ln | {\bf
    M}(t;\mbox{\boldmath $\delta$}(0)) {\bf u}_{l} |  \nonumber \\
            & = &  \lim_{t \to \infty} \frac{1}{2 t} \ln |
    {\bf u}^{\dag}_{l} \cdot {\bf H}(t;\mbox{\boldmath $\delta$}(0))
    \cdot {\bf u}_{l} |
\label{36}
\end{eqnarray}
exist for mixing systems and are independent of the initial conditions.
In this equation
$ {\bf u}_{l}$ denote the $L$ orthonormal eigenvectors of the
real and symmetric matrix $  {\bf H}(t;\mbox{\boldmath $\delta$}(0)) =
{\bf M}(t;\mbox{\boldmath $\delta$}(0))^{\dag}
{\bf M}(t;\mbox{\boldmath $\delta$}(0)) $,
where $\dag$ means transpose. They may be taken as the
basis for an arbitrary inital vector $\mbox{\boldmath $\delta$}(0)$, for which
the  long-time behavior  is ultimately determined by that vector component
$\mbox{\boldmath $\delta$}(0) \cdot {\bf u}_{l}$
with the largest associated Lyapunov exponent.

    The classical algorithm for the calculation of the complete
spectrum of Lyapunov exponents due to Benettin
et. al. \cite{ben1,ben2} and others \cite{shn,wol,erue}
requires the
simultaneous integration of the original reference system (\ref{31}) and
of $L$ sets of the linearized equations (\ref{34}) for $L$ initial
tangent vectors $\mbox{\boldmath $\delta$}_{l}(0)$
taken to be orthonormal. However, due to the stretching and folding
operations of the phase flow, these vectors will not stay orthonormal
for $t > 0$ but will be stretched and  rotatad into the direction
of the largest phase-space expansion corresponding to the largest
exponent, and eventually diverge. This is prevented
by reorthonormalizing the vectors after every few time steps.
The Lyapunov exponents are determined from the time-averaged
contraction/expansion factors for the vector norms.

    A conceptual refinement of this algorithm
has been first  proposed by Hoover et. al. \cite{hppl85,phpr88},
and independently by Goldhirsch et. al. \cite{gso}. It has again been
reinvented since then \cite{plagiat}. In this method the vectors
$\mbox{\boldmath $\delta$}_{l}$ are constrained to remain orthonormal for
all times $t > 0$ by the introduction of constraining forces
added to the right-hand side of the linearized motion equations:

\begin{equation}
\dot{\mbox{\boldmath $\delta$}_{l}}(t) =
      {\bf D}({\bf \Gamma}(t)) \mbox{\boldmath $\delta$}_{l}
 - \sum_{l'=1}^{l}  \lambda_{l'l} \mbox{\boldmath $\delta$}_{l'}
\label{37}
\end{equation}

The $\lambda_{l',l}$ are time dependent Lagrange multipliers
and are determined from the orthonormality conditions
$( \mbox{\boldmath $\delta$}_{l} \cdot \mbox{\boldmath $\delta$}_{l'})
     = \delta_{ll'}$:
\begin{equation}
 \lambda_{l'l} =  \left\{
   \begin{array}{ll}
       \mbox{\boldmath $\delta$}_{l}^{\dag} \cdot {\bf D} \cdot
               \mbox{\boldmath $\delta$}_{l} & \mbox{, if  $l' = l$} \\
       \mbox{\boldmath $\delta$}_{l}^{\dag} \cdot {\bf D} \cdot
               \mbox{\boldmath $\delta$}_{l'} +
       \mbox{\boldmath $\delta$}_{l'}^{\dag} \cdot {\bf D} \cdot
               \mbox{\boldmath $\delta$}_{l} & \mbox{, if  $l' < l$}
   \end{array}
\right.
\label{algor}
\end{equation}
The Lyapunov exponents are given by the time-averaged diagonal multipliers
\begin{equation}
\lambda_{l} = \langle \lambda_{ll}(t) \rangle.
\label{39}
\end{equation}
This algorithm is equivalent to the classical algorithm
with continuous reorthonormalization and yields solutions
for the tangent vectors $\mbox{\boldmath $\delta$}_{l}(t)$
identical to the classical ones, if in the latter case
the vectors are reorthonormalized after every time step.
However, the equations (\ref{algor})
together with (arbitrary) initial conditions may be considered as
the defining equations for the orthonormal vectors
$\mbox{\boldmath $\delta$}_{l}(t)$. They
emphasize another property of tangent-space dynamics
which has been virtually ignored up to now \cite{gso,phh90}, namely
the rotation of the orthonormal tangent vectors.

    We have seen that the solution of Equ. (\ref{algor})
constitutes an orthonormal set
of vectors which continuously change their orientation in tangent space
with instantaneous angular velocities
$\frac{\Delta \Theta_{l}(t)}{\Delta t}$.
The  vectors are forced to remain orthogonal to each other
through the force terms proportional to the off-diagonal Lagrange multipliers in
equation (\ref{36}). If viewed in phase space,
these orthonormal vectors move  with the
state point along the reference trajectory and simultaneously reorient such that
$\mbox{\boldmath $\delta$}_{1}$ always turns toward the direction of fastest
phase-space expansion, $\mbox{\boldmath $\delta$}_{2}$
into a perpendicular direction with second largest growth, and so forth.
As a measure for this unitary rotation
we have computed an averaged angular velocity for each  vector
$\mbox{\boldmath $\delta$}_{l}$ defined  by
\begin{eqnarray}
   \omega_{l} & = & \frac{1}{N_{t}} \sum_{n=1}^{N_{t}}
    \frac{\cos^{-1}(\mbox{\boldmath $\delta$}_{l}(t_{n})
    \cdot \mbox{\boldmath $\delta$}_{l}(t_{n} +
    \Delta t))}{\Delta t}   \nonumber \\
              & = & \frac{1}{N_{t}} \sum_{n=1}^{N_{t}} \frac{|\Delta
            \Theta_{l}(t_{n}) |}{\Delta t},
\label{310}
\end{eqnarray}
where $\Delta \Theta_{l}(t_{n})$ is the angle by which the unit vector
$\mbox{\boldmath $\delta$}_{l}$ reorients during a time step
$\Delta t$ at time $t_{n} = n \Delta t$, and $N_{t}$ is the number
of time steps of the simulation. We refer to  these numbers as
``rotation numbers'', and to their whole set
as the ``rotation spectrum'' \cite{phh90}.Thus, $\omega_{l}$ is the
time-averaged modulus of the angular velocity for the reorienting vector
$\mbox{\boldmath $\delta$}_{l}(t)$ in phase space.

    We stress that - unlike the Lyapunov exponents - the rotation
numbers defined in  (\ref{310}) depend on the metric of the
phase space and of the coordinate system used.
There is no multiplicative ergodic theorem for these
quantities, although they are independent of the choice of the initial
conditions.  In this respect they are on  the same level of theoretical
significance as the instantaneous finite-time Lyapunov exponents,
which also depend on the choice of the coordinate system \cite{hhp90}.
This issue clearly needs further investigation.
In spite of these theoretical restrictions, the  rotation numbers
still convey important information about the phase space
dynamics. For example, for isothermal scans
through order-disorder phase transitions
the rotation numbers increase monotonously with density, whereas the
maximum Lyapunov exponent exhibits a pronounced maximum at the
transition density \cite{phh90,dellago1}. The rotation numbers
are also largest for indices belonging to the smallest exponents
believed to be associated with the stability of collective modes
in a fluid. We therefore speculate that the rotation spectra may prove
useful for establishing a link between the linearized dynamics
in tangent space and more traditional descriptions of systems in terms
of (collective) modes.
Examples for rotation spectra are given in Section IV.
Related numbers have been defined by Ruelle \cite{ruel}
and applied to two-dimensional maps by Lambert et. al. \cite{lambert}.

     It is instructive to view the
dynamics of the $\mbox{\boldmath $\delta$}$-vectors not
only in the whole phase space but also in certain qualitatively
different subspaces associated with special degrees of freedom. Since the
phase space is a product of the center-of-mass configuration space $Q$,
of the respective momentum space $P$, of the angular orientation space
$\Omega$ for the molecular axes, and of the associated angular
momentum space $P_{\Omega}$, also the tangent space is decomposed into
respective subspaces $T Q$, $T P$, $T \Omega$, and $T P_{\Omega}$.
We consider projections of the
$\mbox{\boldmath $\delta$}_{l}$ vectors onto
$T X \in \{T Q, T P, T \Omega, T P_{\Omega}\}$:
\begin{equation}
 \mbox{\boldmath $\delta$}_{X,l} = {\cal P}(T X)\mbox{\boldmath $\delta$}_{l}.
\end{equation}
The projection operator $ {\cal P}(T X)$ may be represented as a diagonal
matrix with
elements ${\cal P}_{\alpha \alpha}(T X)$ equal to unity, if the $\alpha$-axis
of $\mbox{\boldmath $\delta$}_{l}$ belongs to $T X$, and zero otherwise.
We compute the time-averaged squared lengths of these
projected vectors:
\begin{equation}
  \delta_{X,l}^{2} = \langle \mbox{\boldmath $\delta$}_{X,l}
              \cdot  \mbox{\boldmath $\delta$}_{X,l} \rangle
\label{312}
\end{equation}
and refer to them as ``mean-squared $X$-components'' of
$ \mbox{\boldmath $\delta$}_{l}$. Of course, for each $l$ they add up to
unity if summed over $T X$. They are a measure for the probability of a
vector $ \mbox{\boldmath $\delta$}_{l}$ of pointing into a direction
of tangent space belonging to the subspace $T X$. They turn out to be
very helpful for the interpretation of our results. A related
quantity, $  \cos^{2} \alpha^{(X)} \equiv \delta_{X,1}^{2} $, has been
discussed recently by D'Alessendro and Tenenbaum \cite{tene},
who refer to $ \alpha^{(X)} $ as a coherence angle.  It
represents  an effective angle between the subspace $T X$ and the
maximum-expansion subspace.

   We have mentioned already that the constrained orthonormal-vector method
of equations (\ref{37}-\ref{39}) is conceptually very useful. However,
any numerical solution of these equations involves quite extensive
vector-matrix operations. Since the
tangent vectors $\mbox{\boldmath $\delta$}_{l}(t)$ obtained in this
way are identical to the vectors generated by the classical method
with continueous reorthonormalization, equations (\ref{37}-\ref{39})
do not offer an improvement from a numerical point of view.
For our numerical work we have therefore used
the classical method of Benettin in combination with
Gram-Schmidt reorthonormalization after every time step.
To avoid the algebraic complexity we employed a
finite-difference variant of
the classical algorithm:  The tangent vectors in (\ref{32}) and
(\ref{33}) were approximated by finite offset vectors between
two phase-space trajectories corresponding to
a finite $ s = 0.0001 $. As many sets of the
original motion equations (\ref{31}) as the required number
of Lyapunov exponents were integrated for the determination
of the satellite trajectories, and  Gram-Schmidt
reorthonormalization was carried out after every time step.
A fourth order Runge-Kutta algorithm with a reduced time step
$\Delta t = 0.001$ was used for the integration.
Every few time steps the molecular bonds were rescaled to their
precise length to compensate for numerical inaccuracies.
In all runs the trajectories were followed for at least 400
time units.

\section{Results}

    The reduced molecular number density $n^{*} = N\sigma^2 / V$
was chosen large enough to ensure sufficient coupling between
translational and rotational degrees of freedom. Two series of simulation runs
with $n^{*}$ equal to 0.4 and 0.5 were performed. For a given
$n^{*}$ the molecular shape was varied by considering systems with different
molecular bond length $d$. The anisotropy parameter $d/\sigma$
was varied between
0.2 and 1.0, and the corresponding variation of the Lyapunov spectrum
was determined. Since $n^{*}$ does not account for the molecular anisotropy,
we characterize  our model systems  also by
an anisotropy-dependent density parameter
$n^{d} = N \sigma ( \sigma +d) / V$ which is, roughly speaking,
the ratio of the occupied volume to the total.
$n^{d}$ becomes equal to $n^{*}$ for isotropic particles.
Some of our results for the maximum Lyapunov exponent $\lambda_{1}$ and
for the Kolmogorov entropy $h_k$ are summarized in Table 1
for $n^{*}= 0.4$,  and in Table 2 for $n^{*}= 0.5$. Here, the
Kolmogorov entropy $h_{K} = \sum_{l = 1}^{L/2} \lambda_{l}$
is given by the sum of the non-negative exponents. For all these simulations
the kinetic energy per molecule was equal to $2$, $2/3$ for each
translational and rotational degree of freedom.
We monitored this partition of the kinetic
energy among the translational and rotational degrees of freedom throughout
the simulation. We found that equipartition among these degrees of freedom
holds  to within 3\% for the whole range of densities,
bond lengths and temperatures.

    The very different $n^{d}$-dependence
of $\lambda_{1}$ from that of $h_k$ already suggests that the
Lyapunov spectrum is considerably affected by the bond length. This may be
verified from Figure 1 in which all spectra with identical
$n^{*} = 0.4$ but different $d$ are displayed, and from the analogous
Figure 2 for $n^{*} = 0.5$.
Of course, each spectrum consists of discrete points only, which are located
at the nodes of the connecting lines.

The index $l$ along the abscissa
merely numbers the exponents (or degrees of freedom), with
$l=1$ for the maximum exponent, 51 for the smallest positive , and
108 for the most negative exponent. For the construction of
Figs. 1 and 2 only the positive branches (54 exponents) of
the Lyapunov spectra were calculated and displayed. Due to the
Smale-pairing symmetry for symplectic systems mentioned in the
Introduction, the negative branch is obtained by reversing the
sign of the positive branch. 6 of the exponents must vanish due to the
5 constants of the motion  - energy, center of mass, and linear momentum -
and the fact that any perturbation in the direction of the phase flow
adds another  vanishing exponent. Vanishing exponents
and their associated phase space directions have indices
$52 \leq l \leq 57$. However, for short bond lengths the exponent
$\lambda_{52}$ in Figs. 1 and 2 does not vanish exactly. This is an
undesirable consequence of the periodic rescaling of the molecular
bonds, but does not affect the shape of the spectra.

     For a few selected systems containing molecules
with the longest ($d = \sigma$) and the
shortest bonds ($d = 0.2 \sigma$) we computed also the
full spectrum of 108 exponents together with the respective
rotation numbers and mean-squared $X$-components  $\delta_{X,l}^{2}$.
For the longest molecule, $d=\sigma$, full rotation spectra are
depicted in Figure 3 for the  densities $n^{*} = 0.4$ and 0.5.
Mean-squared $X$-components are displayed
in Fig. 4 for $n^{*} = 0.5$, $d=\sigma$, and in Fig. 5 for
$n^{*} = 0.5$, $d=0.2 \sigma$.
These results will be discussed further in the following Section.

    In order to relate these findings to more traditional descriptions
of the dynamics, we calculated also various time correlation functions
referring to translational and rotational
motion of our model system. The normalized velocity autocorrelation functions
for molecular center-of-mass motion,
$C_{vv}(t)=\left\langle
{\bf v}(t) \cdot {\bf v}(0)\right\rangle /
   \left\langle{\bf v}(0) \cdot {\bf v}(0)\right\rangle$,
for the systems with the larger density $n^{*} = 0.5$ are shown in Fig. 6.
Their time integral is related to the  translational
diffusion coefficient. Furthermore, the reorientational
dynamics of linear molecules is usually
discussed in terms of the correlation functions for the
spherical harmonics of rank $l$ of the angles specifying the
orientation of the molecular axis. In three dimensions some
of these correlation functions may be accessible by experiment,
such as infrared absorption ($l=1$), Raman scattering ($l=2$),
and neutron scattering (a weighted sum of $l > 0$ terms).
For our planar model these rotational correlation functions are
written as
\begin{equation}
C_{l}(t) = \langle P_{l}(\cos (\Theta(t)) \rangle ,
\end{equation}
where $\Theta(t) = \eta(t) - \eta(0)$ is the angle  the
axis of a molecule reorients during the time $t$, and
$P_l$ denotes a Legendre polynomial of rank $l$.
In Figure 7 the functions $C_{1}(t) = \langle \cos (\Theta(t)) \rangle $
are depicted for the case $n^{*} = 0.5$ and the whole range of
bond lengths. Similar results have been obtained also for $C_2$ but are
not shown here.

\section{Discussion}

    The computation of $L$ Lyapunov exponents requires the
simultaneous integration of $L(L+1)$ first-order differential
equations. Thus, the number of molecules of a system accessible to
present-day computation is much smaller than one is used to for
traditional simulations of dynamical properties such as correlation
functions or transport coefficients. For that reason our system consists
of only $N = 18$ two-center
molecules in a plane, and the simulation box has a typical side length
of $6 \sigma$ (for  $n^{*} = 0.5$). For such small systems
it is difficult to distinguish between fluid and solid phases,
although already systems of only two disks exhibit a first-order
phase transition \cite{alder}.
Nevertheless, we shall use these expressions to convey the general idea.

    Before entering the discussion we stress again that 6 of the
Lyapunov exponents vanish for reasons given above. They are
located in the middle of the spectrum, $52\leq l \leq 57$, and the
associated phase-space directions belong to the central manifold.
Since Gram-Schmidt reorthonormalization has no ordering effect on the
directions of their
$\mbox{\boldmath $\delta$}_{l}$--vectors the respective rotation numbers
in Fig.~3 and the squared $X$-components in Figs.~4 and 5
have no meaning for $52 \leq l \leq 57$. We have nevertheless included
these points in the Figures to draw attention to this peculiarity.

    An inspection of the center-of-mass velocity autocorrelation functions
in Figure 6 reveals  that backscattering - typical for a dense fluid
or solid - occurs only for the largest bond length. For this system
$C_{1}$  in Fig. 7 does not decay, and  the orientation
of the molecules persists for a long time. We conclude that this state
is a solid, whereas all other systems are fluids. This is
also reflected in the shape of the Lyapunov spectra, which
for $d = 1.0$ and $n^{*} = 0.5$ in Fig. 2
is very similar  to that of a two-dimensional solid
formed with isotropic particles \cite{phpr89}. Also the empirical
power law
\begin{equation}
\lambda_{l}/\lambda_{1} = \left[ (\frac{L}{2} - 2 - l)/(\frac{L}{2} - 3)
      \right]^{\beta}
\label{51}
\end{equation}
for $1\leq l \leq L/2 -2$ works reasonably well with an exponent
$\beta = 3/2$ \cite{phpr89}. $\lambda_{1}$ is the maximum exponent.
As expected, the shape of all other
spectra conforms closely to that of atomic fluids \cite{phpr89},
but a representation in terms of such a power law is not appropriate.

     The maximum Lyapunov exponent $\lambda _{1}$
describes the time evolution of the small perturbation which grows
fastest in phase space. It is a local quantity in the sense that it
depends basically on the fastest dynamical events taking place in the
system, namely collisions, for which the velocities
and angular velocities change sign.
It is a reasonable assumption that the fastest phase-space growth takes place in
the linear and angular momentum subspaces. The mean-squared $X$-components
$\delta_{X,l}^{2}$ introduced in Section 3 support this view.
Simulations for atomic systems have revealed that
$\mbox{\boldmath $\delta$}_{1}$ is located almost fully in the
momentum related subspace ${T P}$ of tangent space. For the
linear molecules studied here
the angular-momentum related subspace $T P_{\Omega}$ of tangent space
turns out to be
the most important. Fig.~4 shows that on the average 69\% of the
squared length of $\mbox{\boldmath $\delta$}_{1}$ for the
elongated molecular case ($d=1$) is contributed by
$T P_{\Omega}$. This number even rises to 96\% for the more freely rotating
case in Fig.~5 ($d=0.2$). We conclude that the main reason for the
instability of the phase space trajectory is due to the anisotropy of
the pair potential, and it is mostly accumulated in the angular momentum
subspace.

    The situation changes completely when we consider
$\mbox{\boldmath $\delta$}_{l}$--vectors for $l>1$ pointing
into less-violently expanding or even compressing phase-space directions.
Figs.~4 and 5 reveal that the linear-momentum subspace becomes dominant
for, say, $20 \leq l \leq 51$, still associated with positive exponents.
In this range the prominent contributions to the Lyapunov spectra comes
from translational modes which one is tempted to associate
with generalized ``hydrodynamical'' modes with small but finite
wave vectors. An approximate representation of these translation-dominated
exponents in terms of the power law (\ref{51}) leads to exponents
$\beta < 1$ similar to that of atomic fluids \cite{phpr89}.
It follows that the positive curvature of the Lyapunov spectra in
Figs.~1 and 2 for the most positive exponents (small $l$) is
essentially due to contributions from rotational degrees of freedom
which make themselves felt more distinctly for small $d$ associated also
with small moments of inertia.

    It is interesting to note that the stable phase-space directions
corresponding to the negative Lyapunov exponents are dominated by
the configurational subspace $Q$ and, to a lesser extent, by the
orientation-angle subspace $\Omega$ (Fig. 4). The significance of
this is not clear to us.

    That the maximum exponents of Figs.~1 and 2 increase with
density $n^{*}$ for fixed $d$ and constant temperature  is
easily explained by the increase of the collision frequency. The relative
maximum  exhibited by the positive exponents of Fig.~2 as a function
of the bond length is less obvious, since an increase of $d$ for
constant $n^{*}$ also increases the collision rate. This maximum
is a consequence of the phase transition eventually leading to
a solid for $d = 1$. In the less-dense case of Fig.~1 this
maximum is expected to occur for bond lengths slightly larger than
$\sigma$. Similar maxima for $\lambda_{1}$ as a function of
density have been observed previously for fluid-to-solid
phase transitions in atomic-particle systems \cite{phh90}.

     The rotation numbers measure the average speed
of rotation of the  $\mbox{\boldmath $\delta$}_{l}$--vectors
in phase space. From simple arguments involving the time-reversal
invariance of the original motion equations (\ref{31})
and of their linearized version (\ref{34}) we expect
that the rotation spectra for symplectic systems are
symmetric such that $\omega_{l} = \omega_{L+1-l}$. This
symmetry was observed for atomic systems \cite{phh90}, and is
also  apparent in Fig.~3 for the linear-molecule case with
$d = 1$. The theoretical significance of these numbers is
still controversial. Also the
$N$-dependence of these numbers needs to be investigated.

     The Kolmogorov entropy $h_K$  is a global measure of the
rate with which information is generated by the dynamics and, hence,
of the disorder in such an equilibrium system.
Our numerical results tabulated in
Tables 1 and 2 confirm the general picture outlined above. For
$n^{*}= 0.5$ this parameter varies only very little with
the molecular anisotropy as long as the system remains fluid (Table 2). It
starts to decrease, when the phase transition is
approached, and becomes significantly smaller for the solid. This transition
takes place near
an anisotropy-dependent density $n^{d}= 0.8$. For the
lower density $n^{*}= 0.4$ a similar transition occurs near
$n^{d}= 0.8$, although $d$ must be increased beyond 1 for a solid to be
observed. This ``transition density'' agrees with the
phase-transition density of an isotropic-particle system, at which
the maximum Lyapunov exponent reaches a maximum \cite{phh90,dellago1}.

\section{Conclusions}

     This work is a first attempt to survey the (in)stability
properties of phase-space trajectories for systems of anisotropic
molecules. A detailed analysis of the Lyapunov spectra as a function
of density and molecular anisotropy makes it possible to
distinguish - at least qualitatively - the contributions from
the center-of-mass motion and of molecular reorientation.
We find that the major contributions to the instability of the
phase space trajectory come from the rotational degrees of freedom and,
in particular, from the  angular momentum variables. Translational
center-of-mass motion is much less destabilizing.
Although,
for practical reasons, the systems contain only a few molecules,
the influence of the fluid-solid phase transition on the
Lyapunov instability and the Kolmogorov entropy is clearly seen.
One might speculate that an even more detailed analysis of the
dynamics of the tangent vectors in the respective subspaces spanned by
the center-of-mass coordinates, the corresponding momenta, the
orientation angles, and the angular velocities may lead to an
interpretation in terms of collective modes in many body systems.
We are still far away from this goal, but it is hoped that
recent progress in the methodology of computing Lyapunov spectra
for systems of hard core particles \cite{dellago2,dellago3,dellago1} will also
be useful for the understanding of anisotropic molecular systems.

\section{Acknowledgements}

     We thank Ch. Dellago and W.G. Hoover for many stimulating
and helpful discussions on this and related subjects. This
work was supported, in part, by a grant to I.~Borzs\'ak by the
Soros Foundation (S-2374/93), in part, by the
Austrian Fonds zur F\"orderung der wissenschaftlichen Forschung,
Grant P9677, and, in part, by the Hungarian OTKA, Grant F7218.
We are also grateful to the staff of the
Computer Center of the University of Vienna for the generous allocation
of computer resources.

\newpage

\begin{center}
{\Large {\bf  Tables:}}
\end{center}
\vspace{10mm}

  \begin{tabular}{|l||l|l|l|l|l|} \hline
   {\bf $d$} & {\bf 0.2} & {\bf 0.33} & {\bf 0.5} & {\bf 0.66} & {\bf 1.0} \\
\hline
    $n^{d}$  & 0.48 & 0.532 & 0.6 & 0.664 & 0.8  \\
    $\lambda_{1}$ & 4.11 & 3.74 & 3.57 & 3.59 & 3.47 \\
    $h_K$ & 111.0 & 114.6 & 117.2 & 120.0 & 111.7  \\ \hline
  \end{tabular}
\begin{description}
\item{Table 1:} Simulation results for a density $n^{*}=0.4$.
   $d$ is the molecular bond length and is given in units of $\sigma$.
   $n^{d}$, the anisotropy-dependent
   density, is defined in the text.  The maximum Lyapunov
   exponent $\lambda_{1}$  and the Kolmogorov entropy  $h_{K}$
   are given in units of $(\epsilon/m \sigma^2)^{1/2}$.
\end{description}
\vspace{10mm}

  \begin{tabular}{|l||l|l|l|l|l|} \hline
   {\bf $d$} & {\bf 0.2} & {\bf 0.33} & {\bf 0.5} & {\bf 0.66} & {\bf 1.0} \\
\hline
    $n^{d}$  & 0.6 & 0.665 & 0.75 & 0.83 & 1.0  \\
    $\lambda_{1}$ & 5.08 & 4.56 & 4.36 & 4.02 & 3.16 \\
    $h_K$ & 134.1 & 135.9 & 133.3 & 115.0 & 65.2  \\ \hline
  \end{tabular}
\begin{description}
\item{Table 2:} Simulation results for a density $n^{*}=0.5$.
   $d$ is the molecular bond length and is given in units of $\sigma$.
   $n^{d}$, the anisotropy-dependent
   density, is defined in the text.  The maximum Lyapunov
   exponent $\lambda_{1}$  and the Kolmogorov entropy  $h_{K}$
   are given in units of $(\epsilon/m \sigma^2)^{1/2}$.
\end{description}

\newpage
\begin{center}
{\Large {\bf  Figure captions:}}
\end{center}

\begin{description}
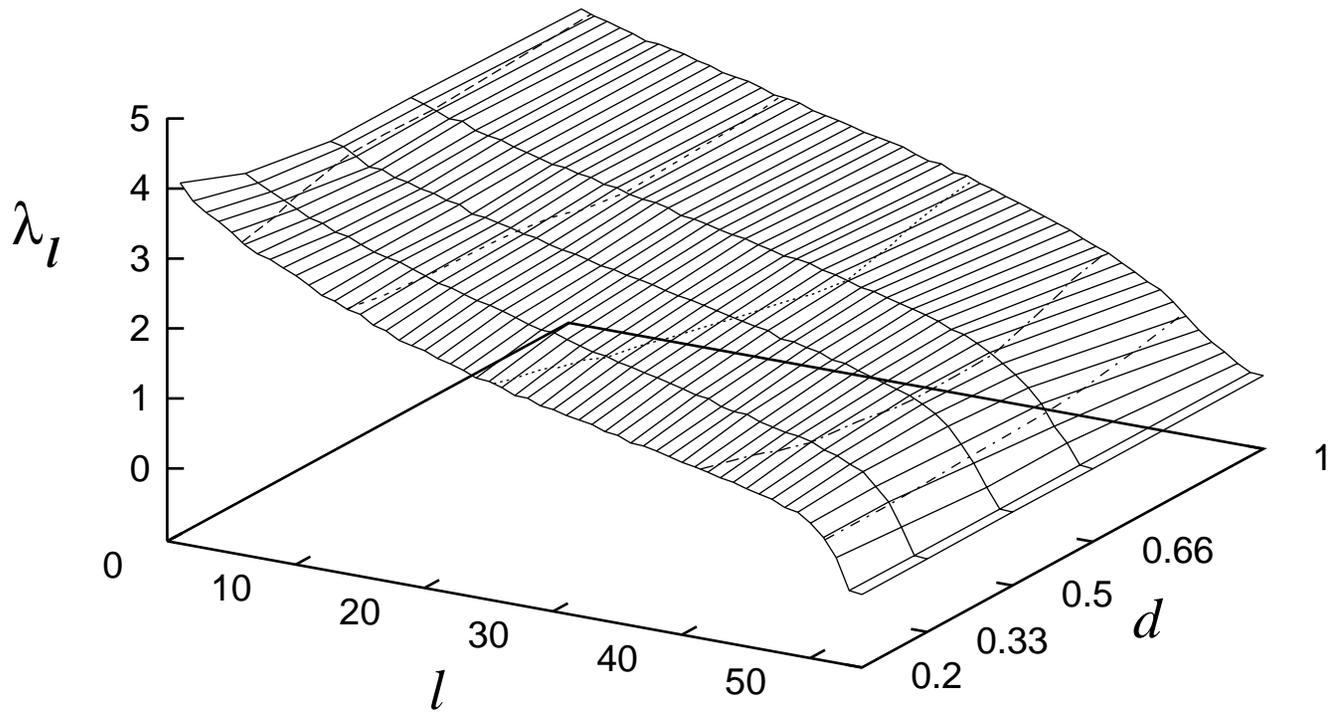
\item[Figure 1:] Lyapunov spectra for a planar fluid of 18
      linear diatomic molecules at a reduced density $n^{*}=0.4$,
      for five different bond lengths $d/\sigma$ equal to
      0.2, 0.33, 0.5, 0.66, and 1.0.  Only the positive branch
      (54 exponents, of which 3 vanish) is shown. The Lyapunov
      exponents are given in units of $(\epsilon/m \sigma^2)^{1/2}$.
      As usual, the index $l$ merely numbers the exponents,
      $\lambda_l$, such that $l=1$ refers to the maximum exponent.

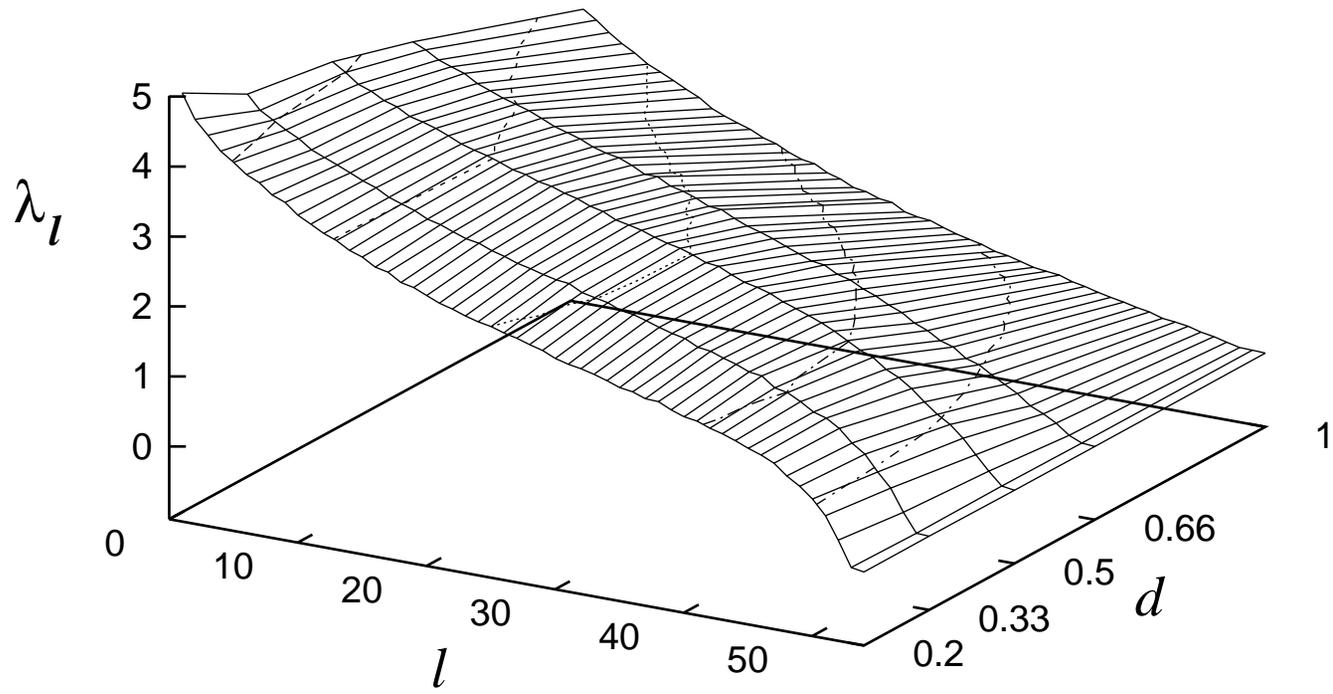
\item[Figure 2:] Lyapunov spectra for a planar fluid of 18
      linear diatomic molecules at a reduced density $n^{*}=0.5$,
      for five different bond lengths $d/\sigma$ equal to
      0.2, 0.33, 0.5, 0.66, and 1.0.  Only the positive branch
      (54 exponents, of which 3 vanish) is shown. The Lyapunov
      exponents are given in units of $(\epsilon/m \sigma^2)^{1/2}$.
      As usual, the index $l$ merely numbers the exponents,
      $\lambda_l$, such that $l=1$ refers to the maximum exponent.

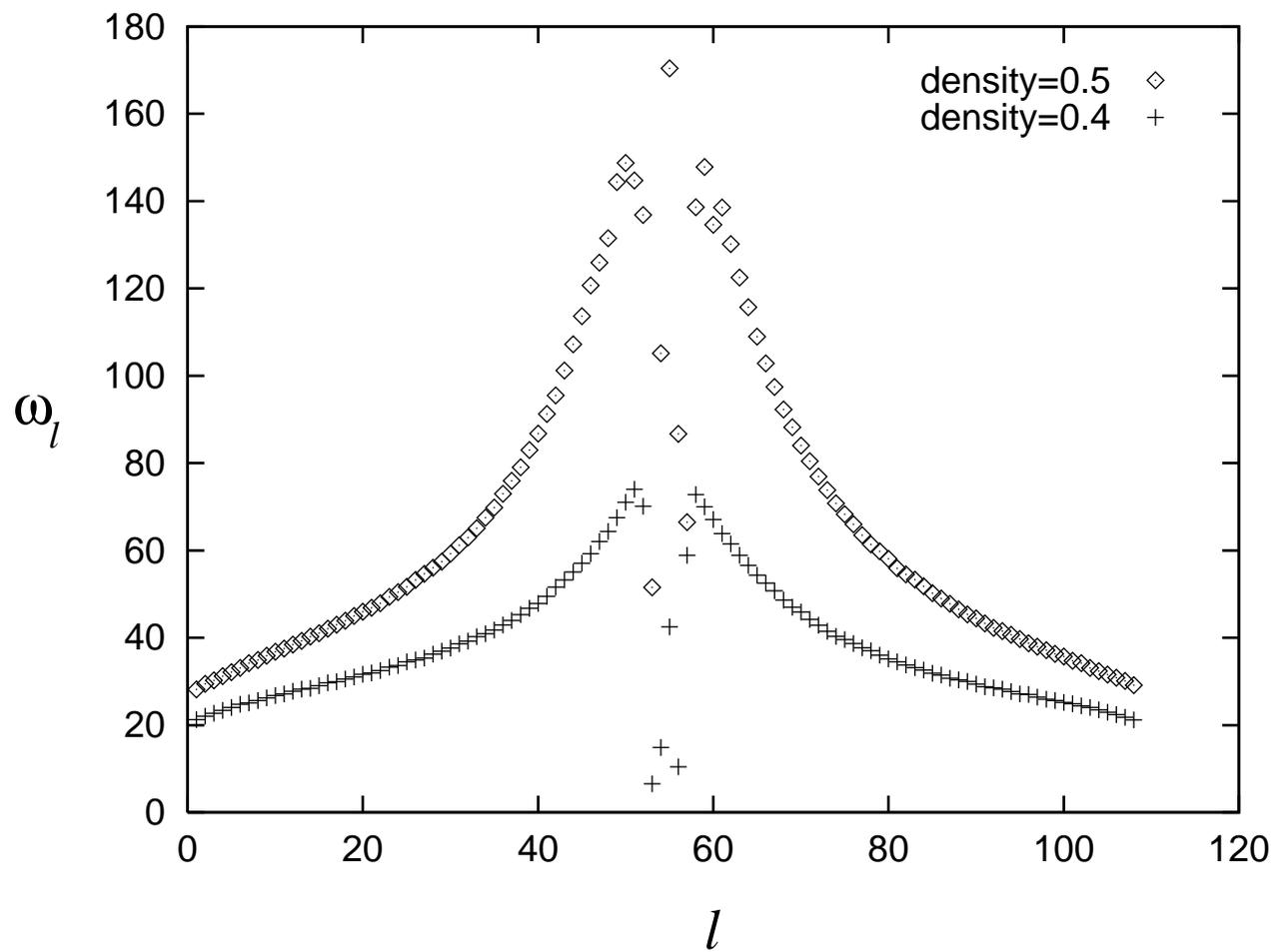
 \item[Figure 3:] Full rotation spectra (108 rotation numbers each) as defined
      in Section 3, for a planar fluid of 18 linear diatomic molecules
      with a bond length $d/\sigma =1$, for the densities $n^{*}$ equal to
      0.4 (plus signs), and 0.5 (diamonds).  The rotation numbers
      are given in units of $(\epsilon/m \sigma^2)^{1/2}$.
      The index $l$ merely numbers the spectral points such that
      1 corresponds to the maximum Lyapunov exponent, 108 to the
      minimum exponent.

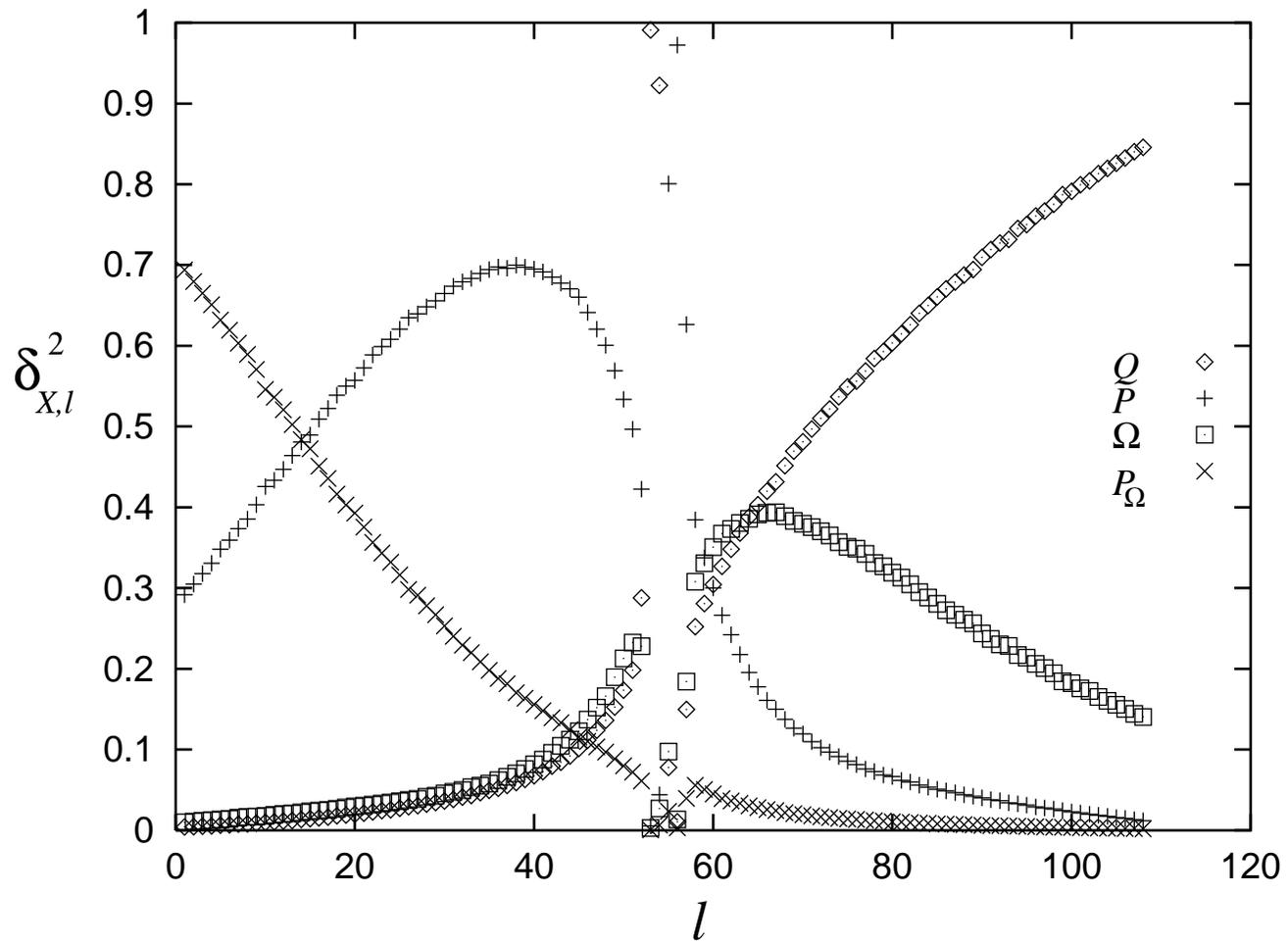
\item[Figure 4:] Mean squared $X$-components   $\delta_{X,l}^{2}$
      as a function of the Lyapunov index $l$ for a planar fluid
      of 18 linear diatomic molecules with a density $n^{*}= 0.4$ and
      bond length $d/\sigma=1$. $l=1$ corresponds to the maximum
      Lyapunov exponent. The subspaces $X$ are the
      center-of-mass configurational subspace ${Q}$ (diamonds), the
      center-of-mass momentum subspace ${P}$ (plus signs), the
      orientation-angle subspace $\Omega$ (squares), and the angular
      velocity subspace ${P}_{\Omega}$ (crosses).

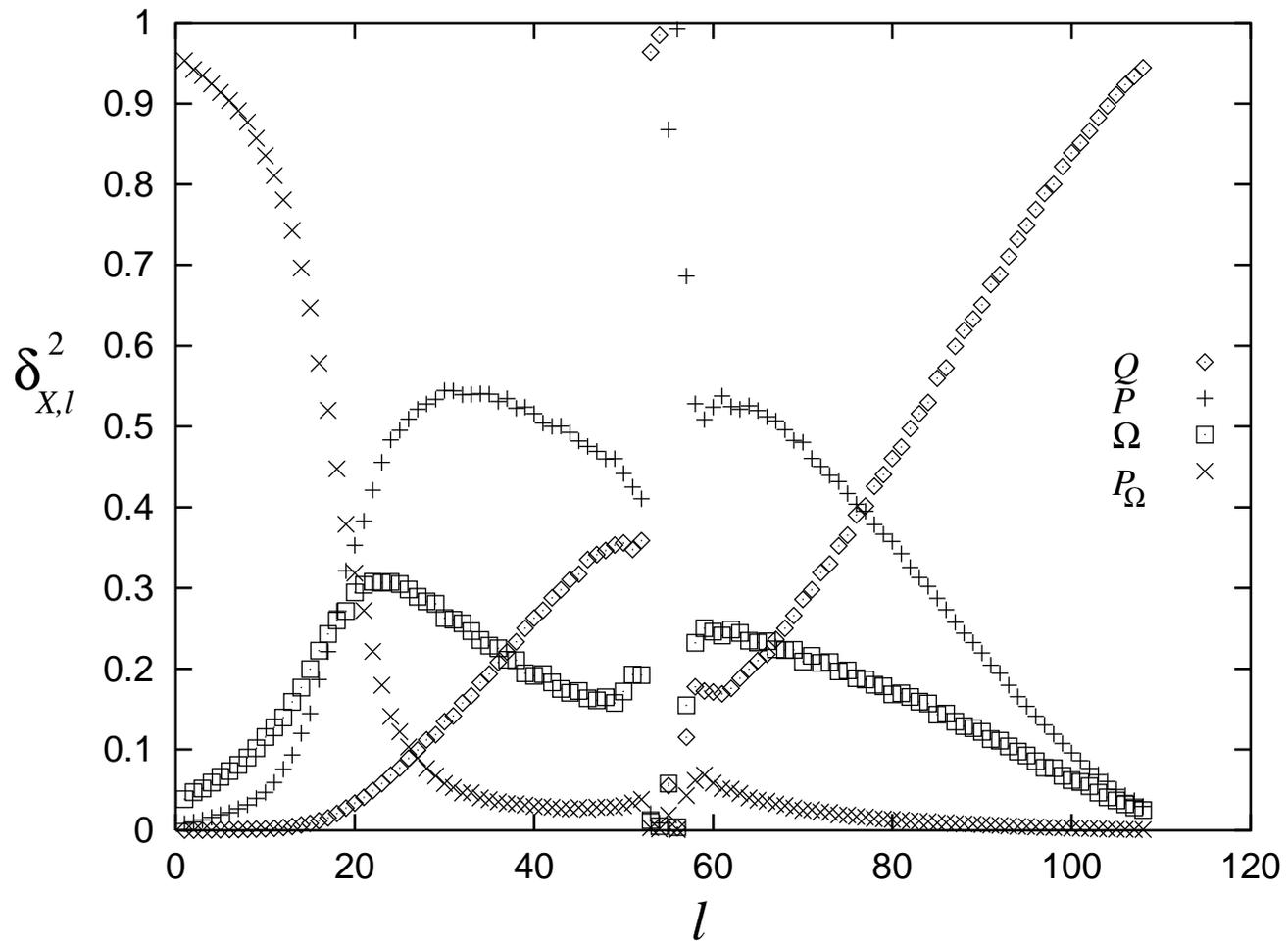
\item[Figure 5:] Mean squared $X$-components   $\delta_{X,l}^{2}$
      as a function of the Lyapunov index $l$ for a planar fluid
      of 18 linear diatomic molecules with a density $n^{*}= 0.4$ and
      bond length $d/\sigma = 0.2$. $l=1$ corresponds to the maximum
      Lyapunov exponent. The subspaces $X$ are the
      center-of-mass configurational subspace ${Q}$ (diamonds), the
      center-of-mass momentum subspace ${P}$ (plus signs), the
      orientation-angle subspace $\Omega$ (squares), and the angular
      velocity subspace ${P}_{\Omega}$ (crosses).

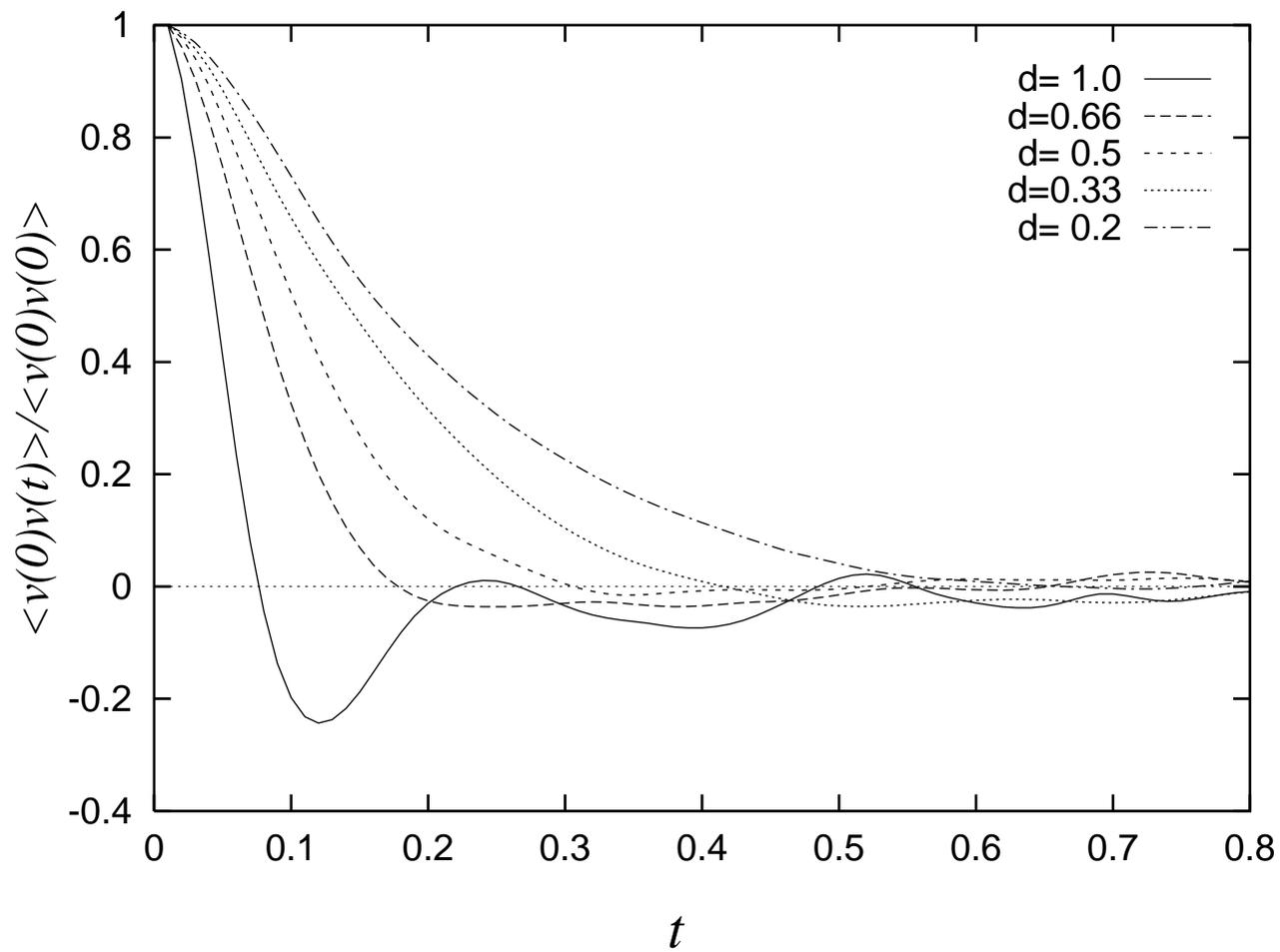
\item[Figure 6:] Normalized velocity autocorrelation functions
      for a system of 18 linear diatomic molecules with a density
      $n^{*}= 0.5$ and different bond lengths $d/\sigma$ varying
      between 0.2 and 1 as indicated. The time is given in
      units of $(m \sigma^2)^{1/2}$.

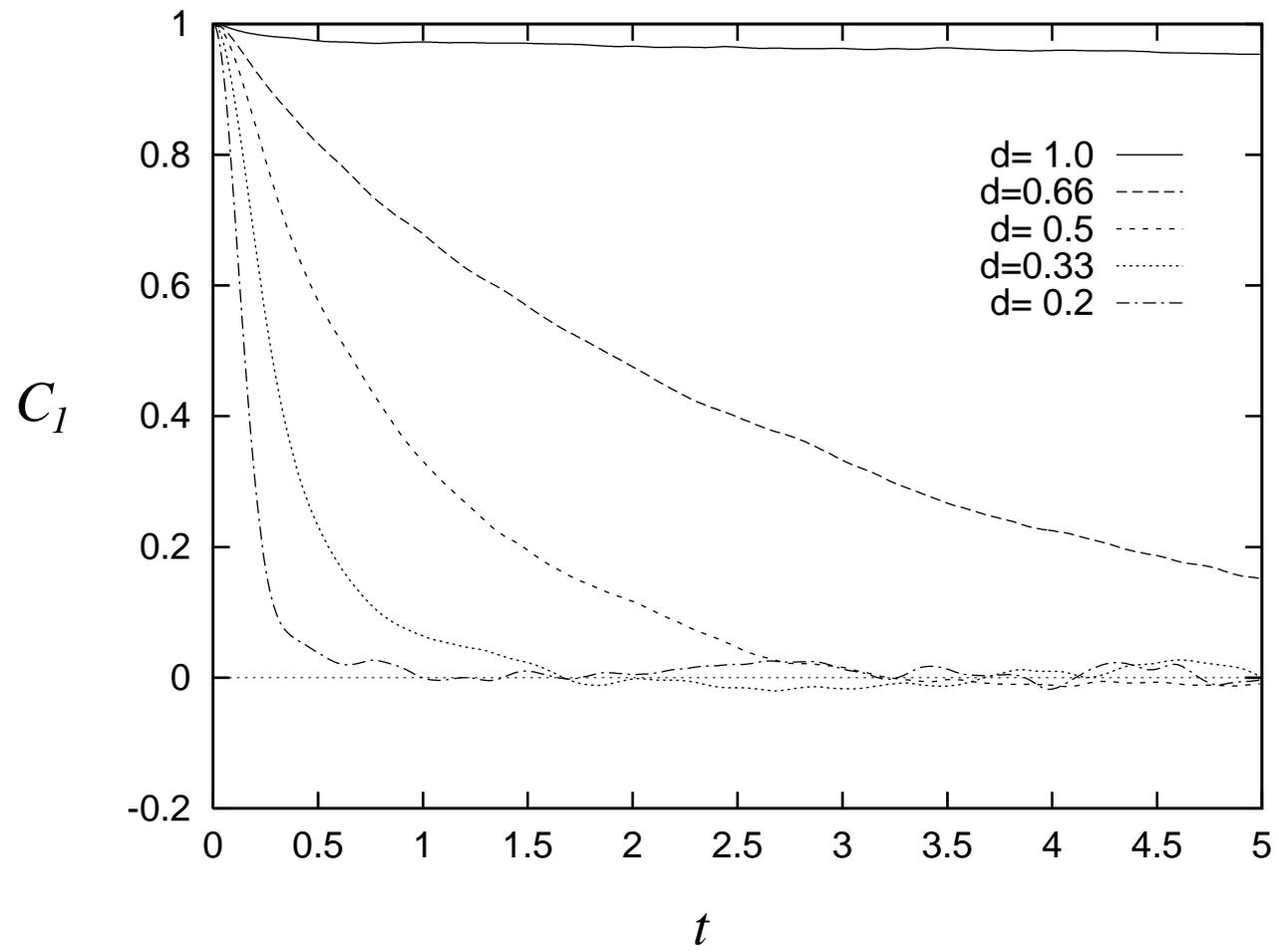
\item[Figure 7:] Orientational correlation functions
      $C_{1}(t) = \langle P_{1}(\cos \Theta(t)) \rangle$
      for a system of 18 linear diatomic molecules with a density
      $n^{*}= 0.5$ and different bond lengths $d/\sigma$ varying
      between 0.2 and 1 as indicated. The time is given in
      units of $(m \sigma^2)^{1/2}$.
\end{description}

\end{document}